\documentclass{article}
\usepackage{psfig}
\begin{document}
\title{Phase transitions in homogeneous
biopolymers: basic concepts and methods.
\footnote{ \uppercase{P}roceedings of the conference
on {\em
`` \uppercase{L}OCALIZATION AND 
ENERGY TRANSFER IN NONLINEAR SYSTEMS"},
\uppercase{J}une 17-21, 2002, \uppercase{S}an 
\uppercase{L}orenzo de \uppercase{E}l \uppercase{E}scorial, \uppercase{M}adrid, 
\uppercase{S}pain. \uppercase{T}o be published by \uppercase{W}orld 
\uppercase{S}cientific.}
} 
\author{N. Theodorakopoulos 
\footnote  {\uppercase {W}ork partially
supported by 
\uppercase {EU} contract \uppercase
{HPRN-CT-1999-00163 (LOCNET} network).    } \\
\\
\it Theoretical and Physical Chemistry Institute,\\
\it National Hellenic Research Foundation,\\
\it Vasileos Constantinou 48, 116 35 Athens, Greece}
\maketitle

{\small The basic models of helix-coil transitions in biomolecules are introduced.
These include phenomenological, zipper (Bragg-Zimm) models of polypeptides,
loop-entropy (Poland-Scheraga) and Hamiltonian (Peyrard-Bishop)
models of homogeneous DNA denaturation. The transfer integral 
approach to one-dimensional thermodynamics is 
presented in some detail, including the
necessary extensions to deal with the singular integral equations arising
in the case of on-site potentials with a flat top. The
(non-)applicability of the theorems which prohibit phase transitions 
in one-dimensional systems is discussed.  }   

\section{Introduction}
The purpose of these notes is to provide a brief introduction to
concepts and methods employed in the description of thermodynamic
phase transitions in model biomolecular systems. This is one of the
areas where the LOCNET network has been active,  with an emphasis
on modelling the fundamental interactions which provide a basis for understanding
both the cooperative behavior and the nonlinear dynamics 
(what I will call ``Hamiltonian'' models). 
My primary aim  is 
therefore to present a working introduction to this area, including some
necessary details on methods and tools;  I hope that young researchers
who enter the field will find this material useful.

On the other hand there is a long and distinguished tradition in the field,
which has achieved remarkable progress, based on a phenomenological
description of the statistical properties of helix formation and growth and
the entropies associated with them.
I have therefore chosen to include
a section on fundamentals of these ``zipper" and ``loop"-based models\cite{PSbook}.  
Again, the hope
is that researchers who are active in developing microscopic Hamiltonian models
will be guided by the succesful features of  phenomenological, ``zipper" and ``loop" models. 
In addition, there is a simple, utilitarian reason for doing this: most of the experimental data
presently available has been analyzed, directly or indirectly, in terms of such models. 
Therefore, anyone seriously interested in comparing theory and experiment should be
familiar with them. 

The plan of this review is as follows: Section  2 deals with zipper and loop models
of polypeptides and DNA, respectively. Section 3 describes the Hamiltonian approach
to DNA denaturation, including a somewhat detailed introduction to the transfer integral
method. An appendix discusses the (non-)applicability of theorems which prohibit 
phase transitions in one dimension to the models presented.

\section{Zippers and Loops}
\subsection{Helix-Coil transitions in polypeptides}
\subsubsection{Background}
Synthetic polypeptides, i.e. macromolecules consisting of identical
amino-acid residues, are ideal for studying the
transition from the alpha-helical to coil-like structure.
Understanding of this transition is central to controlling
the stability of secondary protein structure\cite{Creighton}. 
Residues in helical regions give rise to distinct experimental
signatures (e.g. viscocity, optical rotation).
At a given macromolecular size $N$ - which can be controlled in synthetic polypeptides -
one can measure  the helix fraction as a function of temperature. 
Typically\cite{Doty}, that fraction
completes the transition from 1 to 0 over a fairly narrow 
temperature range -  a few degrees K in the case of long
chains.  Chemists describe the process $A  (helix) \longleftrightarrow   B (coil) $
 as an equilibrium between the two species,
\begin{equation}
K \equiv \frac{c_{B}}{c_{A}} \equiv e^{-\Delta G/T}
\end{equation}
where the helix fraction is given by
\begin{equation}
\Theta \equiv   \frac{c_{A}}{c_{A}+ c_{B}}=\frac{1}{1+K} \>.
\end{equation}
The sign convention is as follows: I am looking at the conversion of
helix (A) to coil (B); therefore $\Delta G = G_{B}-G_{A}
= \Delta H - T \Delta S $,
and $\Delta H > 0$, i.e. the helix is
energetically favored. \par
The value $\Theta = 0.5 $  defines the midpoint of
the transition, $T_{m}$. 
Assuming (although this is not exact, and sometimes not
even a good approximation) that the enthalpy and entropy differences 
do not depend very much on temperature,
leads to
\begin{equation}
\left( \frac{d\Theta }{dT}  \right)_{\Theta =0.5}
= -\frac{1}{4}\frac{\Delta H}{T^{2}} \>.
\label{invwidth}
\end{equation}
\begin{figure}
\psfig{figure=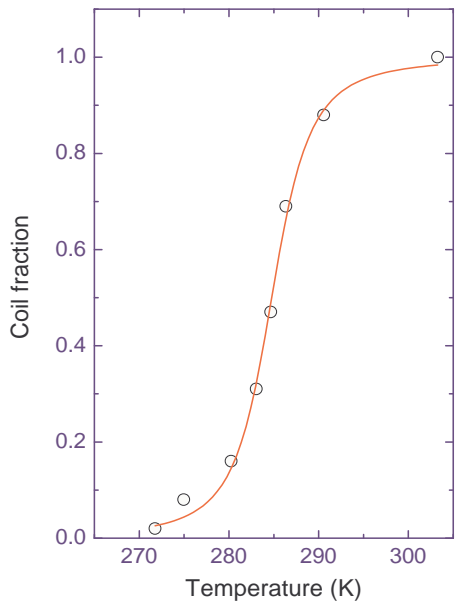,height=9.8truecm,width=7truecm}
\caption { 
Coil fraction vs. temperature for a polypeptide
(poly-$\gamma $-benzyl-L-glutamate) of controlled ($N=1500$) length;
The curve has been calculated in
the framework of the generalized zipper model - cf. Eq. \ref{eq:genziptheta}  
below - 
(redrawn after Ref. {\protect  \cite{Doty}  } ). 
} 
\label{helixcoil}
\end{figure}
The inverse of Eq. (\ref{invwidth})
measures the width of the transition (in degrees K).
A sharp transition (of a few degrees K) has a high
[van't Hoff] $\Delta H$ (of the order of 100 Kcal/mol),
indicating that perhaps as many as 100 hydrogen bonds
are cooperatively broken during the transition.\par
\subsubsection { ``Zipper'' model family: underlying concepts}
\label{subsubsec:hcconcepts}
Helix {\em initiation} and helix {\em growth} are viewed\cite{ZimmBragg} 
as distinct processes:
\begin{itemize}
\item Growth: an existing helix may grow further 
at the $n th$ site, or shrink. This is viewed as a 
forward and reverse reaction, with a rate ratio $s=\exp(-\Delta G^{*}/T)$,
which reflects the difference in local free energies 
between the helix and coil states. If the ratio is greater than unity, 
the helix has a tendency to grow (``zip").   If it is less than unity, the
helix will shrink (``unzip").  At temperatures near the transition,
$s \approx 1$.  The enthalpy difference $\Delta H^{*}<0$ corresponds
to the energy of a single hydrogen bond formed in
the process of helix growth.
\item Nucleation: in order to initiate
a helix, 3 residues have to organize themselves. Again, 
viewing nucleation as a forward / reverse reaction, 
introduces a dimensionless $\sigma = \exp(-\Delta G_{init}/T)$.
The large difference in the free energy comes mostly from
the entropy loss associated with the organization of the 
3-4 residues involved in the first turn of the helix.
\end{itemize}

I now present an outline of theoretical models\cite{ZimmBragg}, in order of 
increasing complexity: 
\subsubsection {``0-th order" - The "all or nothing" (AON) model:} 
Only two states are significant within this model. 
The pure coil, with relative 
statistical weight equal
to unity; and the helix with $N$ residues, with a relative 
weight $\sigma s^{N}$. 
Intermediate states are
suppressed, presumably due to high rate barriers.
This gives a helix fraction
\begin{equation}
\Theta = \frac{1}{N} \frac{N \sigma s^{N}}{1 + \sigma s^{N}}
\end{equation}
and a slope at midpoint
\begin{equation}
\left( \frac{d\Theta }{dT}  \right)_{\Theta =0.5}
=  \frac{N}{4}\frac{\Delta H^{*}}{T_{m}^{2}} \>.
\label{eq:AON}
\end{equation}
There is strong cooperativity.
\subsubsection{ Further considerations: the zipper model}  
The model allows a single connected  helical region of any
length $n\leq N$. The statistical weight (Boltzmann
factor) is -according to
the general considerations, cf. above-, $\sigma s^{n}$, and the helix
can commence at any of the first $A_{n} = N-n+1$ positions. This
gives a partition function
\begin{equation}
Z = 1 + \sum_{n=1}^{N} A_{n} \sigma s^{n}
\end{equation}
and a helical fraction
\begin{equation}
\Theta  = \frac{1}{Z}\sum_{n=1}^{N} n A_{n} \sigma s^{n}
=  \frac{s}{Z} \frac{\partial Z}{\partial s}  \quad,
\end{equation}

where the partition sum can be evaluated to give
\begin{equation}
Z(N) =  1 + \sigma s^{N+2} - (N+1)s + \frac{N}{(s-1)^2} \quad.
\end{equation}

\subsubsection{ The generalized zipper model }  
\label{subsubsec:genzip}
The only difference is topological; then macromolecule consists
of any number of helical and coil
regions which may alternate freely.  
One associates the following
weights:
\begin{itemize}
\item 1 if coil comes after helix or coil;
\item $s$ if helix comes after helix;
\item 
$\sigma s$ if helix comes after coil (nucleation).
\end{itemize}
The model thus implements the ideas presented in section
\ref{subsubsec:hcconcepts}
without imposing any further constraints. The
state of the residue at site $i$ can be described by
a 2-vector $\nu_i$, and the 
partition function is given by
\begin{eqnarray}
\nonumber
Z_N &=& \sum_{ \{ \nu_{1}\}...\{\nu_{N} \} }   < \nu_{1}|T|\nu_{2}>
< \nu_{2}|T|\nu_{3}> ... < \nu_{N-1}|T|\nu_{N}>\\
&=& \sum_{ \{ \nu_{1}\}, \{\nu_{N} \} }   < \nu_{1}|T^N|\nu_{N}>
\end{eqnarray}
where the matrix elements of ${\bf T}$
express the Boltzmann factors specified above, i.e.
\begin{equation}
{\bf T}=
\left(
\begin{array}{cc}
s   & 1  \\
 \sigma s  &   1 
\end{array}
\right)  \>.
\end{equation}
To evaluate the partition sum, I apply periodic boundary conditions 
(convenient, not a must, and certainly wrong for small chains) and
obtain
\begin{equation}
Z_N=Tr {\bf T}^N = \lambda_{0}^{N} +  \lambda_{1}^{N}
\end{equation}
where the eigenvalues are given by 
\begin{eqnarray}
\lambda_{0,1} &=& \frac{1}{2} \left[ 1 + s \pm \Delta \right]\\
\nonumber
\Delta &=& \sqrt{(1-s)^2+4\sigma s}
\label{ziproots}
\end{eqnarray}
and, in the large $N$ limit, $Z$ is dominated by the largest eigenvalue,
$\lambda_{0} $. \par
Note that the partition function (not the {\bf T} -matrix!) can be mapped onto
the one of the ferromagnetic Ising model with exchange interaction \emph{J} and
magnetic field \emph{h}, with the identifications
\begin{eqnarray}
s &\Leftrightarrow & e^{-2\beta h}\\
\sigma  &\Leftrightarrow & e^{-2\beta J}\\
\lambda_{magnetic} &=& e^{\beta (J+h)} \lambda_{helix-coil} \>.
\end{eqnarray}
To obtain the helix fraction, note that if the probability of obtaining
a helical segment of length $k$ is given by  $\phi_k(\sigma)k s^k$, where
$\phi$ is the coefficient of $s^k$ in the partition sum. This gives
\begin{equation}
\Theta = 
 \frac{1}{N}\frac{1}{Z}\sum_{k=1}^{N} \phi_k(\sigma)k s^k     
=\frac{1}{N}\frac{s}{Z}  \frac{\partial Z}{\partial s} 
\label{eq:genziptheta}
\end{equation}
One can now verify that as $s\to 1$, $\Delta \to 2\sqrt{\sigma},
\Theta \to 1/2$; for $\sigma<<1$ (cf. below), 
\begin{equation}
\left( \frac{d\Theta }{dT}  \right)_{\Theta =0.5}
= \frac{1}{4\sqrt{\sigma}}\frac{\Delta H^{*}}{T_{m}^{2}} \>.
\end{equation}
The experimental situation\cite{Doty} for 
long ($N=1500$) chains
is summarized in Fig.  \ref{helixcoil}. Fits can be obtained  
with  $\Delta H^{*}=-3.8 kJ/mole$ (cf. calorimetric
measurements $\Delta H=-3.97 kJ/mole$) and $\sigma=1.6\times 10^{-4}$.  
One usually interprets $1/\sqrt{\sigma}$ as number of residues cooperatively
involved in the transition (cf.  AON theory, Eq. \ref{eq:AON}). This interpretation also 
follows from the Ising model, where the inverse correlation length
is given (in lattice constants) by  
\begin{equation}
1/\xi=
\lambda_{1}-\lambda_{0}
= 2 \sqrt{\sigma} \> \mbox{ (at $s=1$)  } .
\end{equation}

The following simple conformational argument provides an 
independent estimate for 
$\sigma=e^{-\Delta G_{init}/T} \approx e^{\Delta S_{init}} $.
Initiation of the helix involves organization of $J$ (=3 or 4) 
residues, each one by 2 dihedral angles. Typically a dihedral
angle can take 3 independent orientations in space. This
gives a total of $3^{2J}$ states, or an entropy loss
$\Delta S_{init}=-2J \ln3$. (=-6.6 for $J=3$, or -8.8 for $J=4$). This
compares favorably with $\ln(2 \times 10^{-4 })=-8.5$.\par
Similarly, one can relate the entropy loss involved
in helix {\em growth}, to the energy of the H-bond. 
At the transition, $\Delta S^{*}=\Delta H^{*}/T_{m}=-1.85$. This
is roughly comparable to the estimate $-2\ln 3 \approx -2.20$, obtained by
considering the 2 dihedral angles which must be organized to
admit a residue into the helix.\par
\subsubsection{A useful shortcut}
\label{subsubsec:shortcut}
It is possible to obtain the thermodynamics of the generalized zipper
model without recourse to the transfer matrix formalism. I present this
``handwaving" \cite{Azbel}, because it will be useful for DNA loops 
(cf. below).

The fundamental ``entity'' of the macromolecule is a helical region of
length $n$, followed by a coil region of length $m$.  This ``helix-coil" entity is 
characterized by a free energy 
\begin{equation}
g(n,m)= -T\ln \sigma -n T \ln s 
\label{fehc}
\end{equation}
where the two terms correspond to the contributions
of helix nucleation and growth, respectively; it 
occurs
with a probability 
\begin{equation}
P_{n,m}= \exp \{-[g(n,m)-g_0]/T\}
\label{prob_ent}
\end{equation}
where $g_0 \equiv -T\ln z $ is the equilibrium 
free energy per site of the full macromolecule, to be determined by
the normalization condition 
\begin{equation}
\sum_{n,m=1}^{\infty } P_{n,m} = 1  \quad .
\label{pnorm}
\end{equation}
Both the $n$ and the $m$- summations can be done trivially as long as 
$s<z$ and $z>1$.  
The condition (\ref{pnorm}) can then be written as
\begin{equation}
\frac{s}{z-s }\frac{ 1}{z-s }=\frac{ 1}{\sigma  } \quad,
\label{pnormii }
\end{equation}
whose roots are identical to those of (\ref{ziproots}) obtained via
the transfer matrix; the largest root
is the one which satisfies the condition $z>\max(1,s)$ (cf. above). 

\subsection{Loop entropies and DNA denaturation}
\subsubsection{Background}
\label{subsubsec:DNAback}
Thermal DNA denaturation occurs when the two strands of the double helix
separate upon heating. In real DNA the phenomenon of multistep melting is
ubiquitous, reflecting the inhomogeneity of the molecule. 
``Homogeneous", synthetic
DNA, which consists of a few thousand 
identical base pairs has been studied experimentally\cite{synthetic}
and shown to exhibit a very sharp transition; the qualitative
shape of the coil fraction curve, as obtained by optical density
or viscocity measurements, is similar to that of Fig. \ref{helixcoil};
however, the observed temperature width is of the order of one degree. 
It is reasonable to 
speculate that in the thermodynamic limit the transition would be 
of the first order.
Poland and Scheraga\cite{PoSche}(PS)  proposed a simple model of
the thermodynamics involved, based on the ideas discussed in 
Section \ref{subsubsec:hcconcepts}, and the concept of loop entropy (cf. below). 
There is however a difference
in the physical origin of the parameters involved; the helix growth probability
$s=\exp\{-\epsilon  /T \}$ now reflects the combined effect of both 
interactions which are significant in the bonded DNA state: the hydrogen bonding 
responsible for binding base pairs and the stacking interaction 
between adjacent bases. 
\subsubsection{Loop entropies}
\label{subsubsec:loopentropies}
A denaturation loop of length {\em m}, i.e. a region of {\em m} sites 
where the double helix has locally melted,
is characterized by the extra entropy it contributes\cite{PoSche}. This extra
entropy has been calculated for lattice polymers and is of the form
\begin{equation}
S_{L}(m) = am + b - c \ln m  \quad,
\label{loopentropy}
\end{equation}
where $a$ and $b$ are constants and $c$ depends on the dimensionality.
In the case of Gaussian polymer chains, or random lattice walks, which 
ignore the effects of excluded volume, $c=d/2$.  Taking account of 
excluded volume tends to increase the value of $c$.  It will be seen below 
that this can have a decisive influence on the nature of the transition. 

\subsubsection{The phase transition}
In an infinitely long DNA 
molecule, the fundamental entity is again a
double-helical region of $n$
sites (base pairs), followed by a denaturation loop of length $m$
($2m$ bases); putting together the contributions from helical and
loop part (cf. Sections \ref{subsubsec:DNAback} and \ref{subsubsec:loopentropies}   ), 
I obtain the free energy of this entity as 
\begin{equation}
g(n,m) = - T\ln \sigma + n \epsilon -m T \ln u^{2}  + T c \ln m 
\label{fehl}
\end{equation}
where $u=\exp(a/2)$ is another constant, and I have dropped the
irrelevant constant $b$. It is now possible to derive the the thermodynamics
exactly as in section \ref{subsubsec:shortcut}. Inserting (\ref{fehl}) in the normalization
condition (\ref{pnorm}) gives 
\begin{equation}
\frac{ z}{s }-1=\sigma  U\left(\frac{z}{u^{2}}\right)  \quad,
\label{dnathermo}
\end{equation}
where $s=\exp(-\epsilon /T)$ is the only temperature dependent parameter, and
\begin{equation}
U(x)=\sum_{m=1}^{\infty }\frac{ 1}{ m^{c }} \left( \frac{ 1}{ x}\right)^{m }
             \quad.
\label{Uofx}
\end{equation}
Near $x=1$ it is possible to approximate $U(x)$ by the expression\cite{Fisher66}
\begin{equation}
U(x) \approx U_{0} - U_{1}\left[ 1-\frac{ 1}{x }   \right]^{c-1 }
\label{Uapprox}
\end{equation}
where, for $c>1$, $U_0 \approx U_1 \approx \zeta (c)$ \cite{PSbook}.  

I now follow the thermodynamic behavior near the putative singularity by 
defining\cite{Fisher66}
an $s_{c}\equiv \exp(-\epsilon /T_{c})$ via
\begin{equation}
\frac{u^{2}}{s_{ c} }-1 = \sigma U(1)
\label{attc}
\end{equation}
and subtracting (\ref{attc}) from (\ref{dnathermo})
to obtain 
\begin{equation}
s_{c }-s  + \Delta + \frac{ \sigma s_{ c}}{u^{2 } }U_{ 1}\Delta ^{c-1 }=0
\quad, 
\label{thcrit}
\end{equation}
where $z=u^{2}(1+\Delta )$ and I have only kept lowest order terms in
the small quantities
$\Delta $ and $s-s_{c}$. It is now straightforward to use (\ref{thcrit})
and obtain $\Delta (s)$; the helix fraction is then given by
\begin{equation}
\frac{ \partial \ln z}{\partial \ln s } \propto \frac{ \partial \Delta }{\partial s  } 
\quad.
\label{ }
\end{equation}

Two cases can be distinguished:
\begin{itemize}
\item $1<c<2$. The linear term in $\Delta $ can be neglected in (\ref{thcrit}). The helix fraction
is proportional to $(T_{ c}-T)^{(2-c)/(c-1) }$, i.e. it approaches zero continuously near
the transition. In particular,
if $c=3/2$ (the value which corresponds to $d=3$ and neglecting excluded volume effects),
one obtains a second order transition.
\item $c>2$. The linear term in $\Delta $ dominates, and the transition becomes first order,
i.e. the helix fraction drops abruptly to zero at the transition.
\end{itemize}
The above analysis shows how crucial the value of $c$ is in determining the nature 
of the transition. It has been long known\cite{Fisher66} that excluded volume effects, as
calculated within the framework of self-avoiding walks, can increase the value of $c$ to
$1.75$ for loops embedded in three-dimensional space. 
Recent research\cite{Muka} suggests that $c$ may be as high as $2.1$. 
\section {Hamiltonian approach to DNA denaturation}
\subsection{The model} 
A   ``minimal" Hamiltonian model of homogeneous DNA denaturation
has been proposed by Peyrard and Bishop\cite{PBI}(PB). The model
assumes 
two parallel, harmonic  chains, with lattice constant $l$,
joined in the form of a ladder
by anharmonic springs; the particular model proposes
a Morse potential because of its analytical tractability, 
although any form with a repulsive
core, a stable minimum and a flat top (e.g. Lennard-Jones)
would be physically suitable. 
The emphasis is on modelling the unbinding of the
two chains, not the helical aspect of the ordered state; this
is done in the general spirit of 
the theory of critical phenomena, which has demonstrated
that the ``essentials" of
the interactions completely determine the critical behavior. 
The Hamiltonian
\begin{eqnarray}
\nonumber
H_{tot} & = & \frac{m}{2} \sum_{ n}\left[\dot{u}_{n}^{2}+\dot{v}_{n}^{2} +
\omega_{0}^{2}\left(u_{n}-u_{n-1} \right)^{2} 
+\omega_{0}^{2}\left(v_{n}-v_{n-1} \right)^{2}  \right] \\
& + & \sum_{n}V(u_{n}-v_{n})
\end{eqnarray}
describes the motion of the two bound chains with 
coordinates $\{u_{n} \}, \{u_{n} \} $; dots denote time derivatives; 
only the motion transverse 
to the chains is considered; bases
have equal masses $m$ and are connected
by harmonic springs of equal strength, determined by
the frequency $\omega_{0}$; the energy scale of the Morse potential
\begin{equation}
V(x)=D(e^{- a x}-1)^{2} 
\end{equation}
is given by $D$ and its spatial range by $1/a$.

Transformation  to center-of-mass and relative coordinates, 
$Y_{n}=(u_{n}+v_{n})/2, 
y_{n}= u_{n}-v_{n}, M=2m, 1/\mu=2/m$ 
decouples center-of-mass from relative motion, i.e.
\begin{equation}
H_{tot}=H_{0}(Y)+H(y)  \quad,
\end{equation}
where
\begin{equation}
H_{0}(Y)= \sum_{n}\left[\frac{P_{n}^{2}}{2M} 
+\frac{1}{2}M\omega_{0}^{2}(Y_{n}-Y_{n-1})^{2}\right] \quad,       
\end{equation}
$P_{n}=M\dot{Y}_{n}$ is the canonical momentum conjugate
to $Y_{n}$, and
\begin{equation}
H(y)= \sum_{n}\left[\frac{ p_{n}^{2}}{2\mu}
+\frac{1}{2}\mu\omega_{0}^{2}(y_{n}-y_{n-1})^{2}
+V(y_{n})
\label{eq:trham}
\right] \quad,
\end{equation}
where $p_{n}=\mu \dot{y}_{n}$ is the canonical momentum conjugate
to $y_{n}$. \par
\subsection{Statistical Mechanics}
$H_{0}$ is just the Hamiltonian of a harmonic chain
with the total base pair mass $2m$ per site. It gives an additive nonsingular
contribution to all thermal properties. It will be neglected  
in what follows. The classical thermodynamics of $H$ is described by
the canonical partition function 
\begin{equation}
Z = \int \prod _{n=1}^{N }dp_{n}dy_{n} e^{-\beta H} \quad.
\end{equation}
One can immediately perform the Gaussian integrals over
momentum space and obtain
\begin{equation}
Z=Z_{K}Z_{P},
\label{Z}
\end{equation} 
where each integration in the kinetic part
contributes a $(2\pi\mu/\beta)^{1/2}$ factor to the partition function,
i.e.
\begin{equation}
Z_{K}= (2\pi\mu/\beta)^{N/2}  \quad.
\label{ZK}
\end{equation}
 The nontrivial part is
\begin{equation}
Z_{P}= \int \left( \prod_{n=1}^{N}dy_{n} \right)
\>  T\left( y_{1},y_{2} \right) \cdots T\left( y_{N-1},y_{N} \right)
T\left( y_{N},y_{N+1} \right) 
\quad,
\label{ZP}
\end{equation}where
\begin{equation}
T(x,y) = e^{-\beta \left[
\frac{\mu\omega_{0}^{2}}{2}(y-x)^{2}+V(x)   \right]} 
\quad.
\label{kernel}
\end{equation}
\subsection{Transfer integral: the formalism}
\subsubsection{Definitions and Notation}
\label{subsubsec:TIdefnot}
Consider the eigenvalue problem defined by the
asymmetric kernel $T$ (the kernel can be easily
symmetrized but need not be so; in fact, working
with the asymmetric kernel is technically advantageous
in examining the validity of some approximations,
cf. below):
\begin{eqnarray}
\label{right}
\int_{ -\infty }^{ \infty } dy \> T(x,y) \> \Phi_{\nu}^{R}(y) & = &
\Lambda_{ \nu}    \Phi_{\nu}^{R}(x)\\
\label{left}
\int_{ -\infty }^{ \infty } dy \> T(y,x) \> \Phi_{\nu}^{L}(y) & = &
\Lambda_{ \nu}    \Phi_{\nu}^{L}(x)  \>,
\end{eqnarray}
where left and right eigenstates have been assumed to be
normalized; note that the normalization
integral is $\int dx \Phi_{\nu}^{L}(x) \Phi_{\nu}^{R}(x)$.
Orthogonality 
\begin{equation}
\int_{ -\infty }^{ \infty }  dx \> \Phi_{\nu}^{L}(x) 
\> \Phi_{\nu'}^{R}(x) = \delta _{ \nu \nu'} 
\end{equation}
and completeness
\begin{equation}
\sum_{\nu } \Phi_{\nu}^{L}(x) 
\> \Phi_{\nu}^{R}(y) = \delta (x-y)
\label{complete}
\end{equation}
relationships are assumed to hold. Note that this is not obvious for
the class of potentials of interest here. This is a point which will be
further taken in section \ref{subsec:TIbeyondgp}. I will further use the notation
\begin{equation}
\Lambda _{ \nu } = e^{-\beta \epsilon _{ \nu } } 
\end{equation}
(sensible as long as the eigenvalues are nonnegative).

\subsubsection{The partition function}

The integrand of 
(\ref{ZP}), as written down has a problem: it includes
a reference to the displacement $y_{ N+1}$ of the
$N+1 st$ particle, which has not yet been defined.
For a large system, this is best remedied by means
of periodic boundary conditions (PBC), i.e. by demanding
that $y_{ N+1}=y_{ 1}$. Alternatively, 
the integration may be extended to one more variable,  $dy_{ N+1}$,
with the simultaneous introduction of a  factor $\delta (y_{ N+1}-y_{ 1})$ to take
care of PBC.  This however is the same as the sum 
in the left-hand-side of (\ref{complete}).  I then obtain
\begin{equation}
Z_{ P}= \sum_{ \nu } \int dy_{ 1}\cdots 
\underbrace{dy_{ N+1}
\Phi _{ \nu }^{L}(  y_{1}) }
T(y_{ 1},y_{ 2})\cdots 
\underbrace{T(y_{ N},y_{ N+1})
\Phi _{ \nu }^{R}(  y_{ N+1})   }  \>. 
\end{equation}
The braces make clear that I can perform the integral
over $d y_{ N+1}$ and obtain  a factor
$\Lambda _{ \nu } \Phi _{ \nu }^{R}(  y_{ N+1}) $, using the defining property
of right-hand eigenfunctions. The process can be repeated $N$ times,
each time giving a further factor $\Lambda _{ \nu } $ and a right eigenfunction
with an argument whose index is smaller by one. At the end, I am  left
with
\begin{equation}
Z_{ P} = \sum_{ \nu } \int dy_{ 1}
\Phi _{ \nu }^{L}(  y_{1}) \Lambda_{ \nu }^{N }
\Phi _{ \nu }^{R}(  y_{1})\\
  = \sum_{ \nu }  \Lambda_{ \nu }^{N} \>.
\end{equation}
In the thermodynamic limit, $Z_{ P}$ is dominated by the largest eigenvalue
$\Lambda _{ 0}$  or, equivalently, the lowest $\epsilon _{0 }$:
\begin{equation}
\lim_{ N \to \infty } \frac{ 1}{ N}\ln Z_{ P} =  \ln \Lambda _{ 0} =
-\beta \epsilon _{0}
\label{ZPeps}
\end{equation}
\subsubsection{The order parameter} 
\begin{eqnarray}
\nonumber
<y_{ i}>&=&\frac{ 1}{Z_{ P} }\int dy_{1}\cdots dy_{ N}T(y_{ 1},y_{ 2})
\cdots T(y_{ i-1},y_{ i}) y_{ i} 
\\ \nonumber
 &\>&T(y_{ i},y_{ i+1 })\cdots T(y_{ N},y_{ N+1})\\
\nonumber
& \equiv & \frac{ 1}{Z_{ P} } 
\sum_{ \nu } \int dy_{ 1}\cdots 
dy_{ N+1}
\Phi _{ \nu }^{L}(  y_{1}) 
\underbrace{T(y_{ 1},y_{ 2})\cdots T(y_{ i-1},y_{ i})}
_{i-1} y_{ i} 
\\
 &\>& \underbrace{ T(y_{ i},y_{ i+1 })\cdots 
T(y_{ N},y_{ N+1}) }_{N-i+1}
\Phi _{ \nu }^{R}(  y_{ N+1})   \>,
\end{eqnarray}
after insertion of a complete set of states (cf. above); the
braces denote the number of times I can perform an
integration and obtain, respectively,
 a right eigenfunction with an argument 
smaller by one, or a left eigenfunction with an argument
larger by one, as well as a factor $\Lambda _{ \nu }$. 
The remaining integral must be performed explicitly:
\begin{eqnarray}
\nonumber
<y_{ i}>&=& \frac{ 1}{Z_{ P} } \sum_{ \nu } 
\Lambda _{ \nu }^{N }M_{\nu \nu}   \\
& \approx  & M_{00}
\label{OPTI}
\end{eqnarray}
where the second line is exact in the thermodynamic 
limit, and I have used the abbreviation
\begin{equation}
M_{\nu \mu}= \int_{ -\infty } ^{\infty  }dy 
\Phi _{ \nu }^{L}(y) y\Phi _{ \mu }^{R}(y)  \quad.
\end{equation}
\subsubsection{Correlations}
With  $i< j$,
\begin{eqnarray}
\nonumber
<y_{ i}y_{ j}>&\equiv &\frac{ 1}{Z_{ P} }\int dy_{1}\cdots dy_{ N}
T(y_{ 1},y_{ 2})
\cdots T(y_{ i-1},y_{ i}) y_{ i} T(y_{ i},y_{ i+1 })
\\ \nonumber
 &\>&\cdots T(y_{ j-1},y_{ j}) y_{ j} T(y_{ j},y_{ j+1 })
\cdots T(y_{ N},y_{ N+1})\\
\nonumber
& =&  \frac{ 1}{Z_{ P} }
\sum_{ \nu } \int dy_{i}\cdots dy_{ j} \Lambda _{ \nu }^{i-1 }
\Phi _{ \nu }^{L}(  y_{i}) 
y_{ i} T(y_{ i},y_{ i+1 })
\\ 
 &\>&\cdots T(y_{ j-1},y_{ j}) y_{ j}
\Lambda _{ \nu }^{N- j+1 }
\Phi _{ \nu }^{R}(  y_{j}) 
\end{eqnarray}
where the straightforward integrations, i.e the first $i-1$ and the 
last $N-j+1$ have already been performed (cf. above). 
In order to perform the remaining integrations, I insert
two more factors of 1, after $y_{i}$ and before $y_{j}$, i.e.
integrals $\int \delta (y_{i}-{\bar y}_{i}) $
and $\int \delta (y_{j}-{\bar y}_{j}) $, respectively; exploiting the
presence of the $\delta $ functions, I may substitute
the variables $y_{ i}$ and $y_{j}$   by ${\bar y}_{i}$ 
and ${\bar y}_{j}$ respectively. This translates
to two more sums over complete sets of states
and another $j-i$ integrals which can now be performed:
\begin{eqnarray}
\nonumber
<y_{ i}y_{ j}>& =  &  \frac{ 1}{Z_{ P} }
\sum_{ \nu, \mu, \rho    } \int d{\bar y}_{i}d{\bar y}_{j}
dy_{i}\cdots dy_{ j} \Lambda _{ \nu }^{i-1 }
\Phi _{ \nu }^{L}(  {\bar y}_{i}) 
{\bar y_{ i}}     
\Phi _{ \mu }^{R}(  {\bar y}_{i}) 
\Phi _{ \mu }^{L}(   y_{i}) \\ \nonumber
 &\>&  T(y_{ i},y_{ i+1 })
\cdots T(y_{ j-1},y_{ j}) 
\Phi _{\rho}^{R}( y_{j}) 
\Phi _{\rho}^{L}({\bar y}_{j}) 
{\bar y}_{ j}
\Lambda _{ \nu }^{N- j+1 }
\Phi _{ \nu }^{R}( \bar{ y}_{j})  \\
\nonumber
&=&  \frac{ 1}{Z_{ P} }\sum_{ \nu, \mu, \rho} 
\Lambda _{ \nu }^{N+i-j} \Lambda _{ \rho }^{j-i}
\int d{\bar y}_{i}d{\bar y}_{j}
\Phi _{ \nu }^{L}(  {\bar y}_{i}) 
{\bar y_{ i}}     
\Phi _{ \mu }^{R}(  {\bar y}_{i}) 
\\ \nonumber
& \> &
\delta_{ \mu,\rho  } 
\Phi _{\rho}^{L}({\bar y}_{j}) {\bar y_{ j}}     
\Phi _{ \nu }^{R}( \bar{ y}_{j}) \\ 
&=&  \frac{ 1}{Z_{ P} }\sum_{ \nu, \mu} 
\Lambda _{ \nu }^{N+i-j} \Lambda _{\mu}^{j-i}
\left| M_{\nu \mu} \right|^{2 } \>.
\end{eqnarray}
In the thermodynamic limit  the $\nu=0$ term  dominates; the 
resulting factor cancels against the denominator and leaves
\begin{equation}
<y_{ i}y_{ i+r}> = \sum_{\mu}   \left| M_{0\mu}\right|^2
e^{-\beta (\epsilon _{ \mu } -\epsilon _{ 0})r}  
\end{equation}
where I have used Dirac shorthand for the matrix
element and set $j=i+r$. 
The first term ($\mu=0$) in the above sum corresponds to $<y>^2$
and should properly be subtracted from both sides; 
This leaves
\begin{eqnarray}
< \delta y_{ i}\delta y_{ i+r}>& \equiv &
<y_{ i}y_{ i+r}> -<y_{ i}><y_{ i+r}> \\ \nonumber
& = & \sum_{\nu}  \>^{'}   \left| M_{0 \nu} \right|^2
e^{-\beta (\epsilon _{ \nu } -\epsilon _{ 0})r}  
\end{eqnarray}
where now the ground state is excluded from the summation.
The above result identifies the correlation length
$\xi $, i.e  the typical length over which the decay of
correlations takes place, as
\begin{equation}
    \frac{ \xi }{l } = \frac{ 1}{\beta (\epsilon _{ 1 } -\epsilon _{ 0})  }
\label{xiTI}
\end{equation}
where the subscript $1$ stands for the first excited state
(dominant exponential in the limit of large $r$).
\subsection{TI results: Gradient-expansion approximation}
Suppose that the displacement field does not change
appreciably over a lattice constant. This is certainly reasonable
at low temperatures. Note that this does not exclude large
displacements per se. Nonlinearity is explicitly allowed, but
the displacement field must be smooth. The assumption is
certainly reasonable at low temperatures. \par
I set $y=x+z, \Phi^{R} \rightarrow \phi $ and 
rewrite (\ref{right}) as
\begin{eqnarray}
\nonumber
e^{-\beta [\epsilon _{ \nu } - V(x)] }\phi_{ \nu }(x) &=&
\int_{ -\infty}^{+\infty  } dz  \> e^{-\frac{ 1}{2 }\beta \mu \omega _{ 0}^{2 }z^2 }
\left\{ \phi_{ \nu }(x)+ z \phi_{ \nu }'(x)
+\frac{ 1}{ 2} z^{2}\phi_{ \nu }''(x)  \right\}\\
& = &  \left[ \frac{ 2\pi }{ \beta \mu \omega _{ 0}^{2 }}\right]^{1/2}
\left\{ \phi_{ \nu }(x) + 
\frac{ 1}{ 2\beta \mu \omega _{ 0}^{2 } }
\phi_{ \nu }''(x)  \right\}
\label{grexp}
\end{eqnarray}
where higher terms in the gradient expansion have been neglected 
and the Gaussian integrals have been performed; this is meaningful
as long as the width of the Gaussians is smaller than the range of
the Morse potential, i.e. 
\begin{equation}
\beta \mu \omega _{ 0}^{2 }/a^{2} >1  \quad.
\label{sharpGauss}
\end{equation}
The factor in front 
of the r.h.s. of (\ref{grexp}) can be absorbed in the eigenvalue by defining
${\tilde \epsilon }_{ \nu } =  \epsilon _{ \nu } + 
1/(2\beta )]\ln [2\pi/ ( \beta \mu \omega _{ 0}^{2 } )] $.
Now, for many practical
purposes, when it comes to calculating matrix elements,
the relevant magnitude of $\epsilon - V(x)$ 
is $D$, the depth of the Morse well (or some other
characteristic energy in the case of another
potential). The key to this statement is that 
one does not need to consider 
large negative values of $x$, where $V(x)$ is huge, 
because at such $x$, both
the exact eigenfunction $\Phi $ and
its approximation $\phi $ can be expected to
be negligible. 
If then $\beta D\leq 1$\footnote{Note that, in connection with (\ref{sharpGauss}), this
defines a temperature window $  D < k_{B}T <\mu \omega _{ 0}^{2 }/a^{2} $ 
for the validity of the overall approximation scheme.} 
it is reasonable to expand the exponential and
keep only the first term.  
Dividing both
sides by $\beta $, I obtain a Schr\"odinger - like
equation, 
\begin{equation}
-\frac{ 1}{2 \mu (\beta \omega _{ 0}) ^{2 } } \phi_{ \nu }''(x)
+\left[ V(x) - {\tilde \epsilon _{ \nu }} \right] \phi_{ \nu }(x) = 0
\quad.
\label{Schr}
\end{equation}
Before continuing the discussion
of (\ref{Schr}) and its properties, I pick up the
bits and pieces (cf (\ref{Z}), (\ref{ZK}),
(\ref{ZPeps}) ) of the thermodynamic free energy 
  (per site)
\begin{equation}
\label{eq:fsplit}
f =  -\frac{ 1}{\beta N } \ln(Z_{ K}Z_{ P})   \equiv  
-\frac{1}{\beta }  \ln \left( \frac{ 2\pi  }{ \beta \omega _{ 0}}\right)
+   {\tilde  f}   \quad,
\end{equation}
where  ${\tilde  f} =   {\tilde  \epsilon} _{0}$.
The first term in (\ref{eq:fsplit}) is the free energy of the small oscillations
(transverse phonons in this context). It is a term smooth in
temperature (constant specific heat!) and therefore irrelevant to
any phase transition. Any nontrivial
physics is hidden in the second term, which is identical with the the 
smallest eigenvalue of (\ref{Schr}).

A couple of comments are in order.
First, (\ref{Schr}) would be a literal (i.e. quantum-mechanical)
Schr\"odinger equation, if I substituted 
$1/(\beta \omega _{ 0})$ 
by $\hbar$. I will come back to that point.
Second, I can get a dimensionless potential
(and eigenvalue) by dividing both sides of  (\ref{Schr})
by $D$. In other words, the relevant 
dimensionless parameter is
\begin{equation}
\delta^2 = \left\{ \begin{array}{rlr}
 \frac{ 2\mu }{a^{2 } \hbar ^{2 }} \cdot &D    
& \mbox{  (quantum mechanics)   } \\
         &        & \\
\frac{ 2\mu \beta ^{2 } \omega _{ 0}^{2 }}{a^{2 } } \cdot &D 
    & \mbox{(statistical mechanics).} 
\end{array}
\right.
\label{delta}
\end{equation}
In terms of $\delta$, the bound state 
spectrum of (\ref{Schr}) is given \cite{LLQM} by
\begin{eqnarray}
\nonumber
\frac{ {\tilde \epsilon} _{n}}{ D}& = & 
1 - \left[1-\frac{ n+ 1/2}{\delta  }  \right ]^{2} \\
n &= & 0,1, ...  , int(\delta - 1/2)  \>.
\label{boundspectrum}
\end{eqnarray}
There is at least one bound state if $\delta >1/2$. 
For $1\geq \delta > 1/2$  there is  \emph{exactly} one 
bound state. And if $\delta$ becomes equal to, or smaller than
1/2, there is no bound state at all. 
\emph{The value } 
$\delta _{ c}=1/2$ \emph{ is "critical".} 
In quantum mechanical
language, if a particle has a mass which is lighter than a 
critical mass $\mu _{ c}= \hbar^{2 }a^{2 } /(8D)$, it cannot be confined in the
Morse
well. Quantum fluctuations will drive it out\footnote{This is a general
property of asymmetric one-dimensional wells; symmetric wells will support a
particle in a bound state, no matter how low its mass.}.
In the context of statistical mechanics, 
$\delta_{ c}$ corresponds, via (\ref{delta}), to a critical
temperature $T_{ c}=2(\omega _{ 0}/a)\sqrt{2\mu D} $.
The free energy is given by 
\begin{equation}
\frac{{\tilde f} }{D}  = \left\{ \begin{array}{ll}
 1   & \mbox{$T > T_{c}$}\\
  &   \\
 1 -\left( 1 - \frac{T}{T_{ c}}\right) ^{2 }   & \mbox{ $T< T_{ c}$ } \>,
\end{array}
\right.
\label{feMorse}
\end{equation}
where in the upper line I  have made use of the fact that
the bottom of the continuum part of the spectrum is at
$\epsilon =D$. The free energy $f$ is 
non-analytic at $T=T_{c}$, where its  
second derivative is discontinuous (i.e. there
is a jump in the specific heat). This 
corresponds to a second order transition, according to
the Ehrenfest classification scheme\footnote{Note that the term "second
order" is meant literally in this case, not just as a metaphor for
the absence of a latent heat (for which the term "continuous transition"
would be appropriate).}.  

In order to gain some further insight
into the physics involved\footnote{
The mathematical analogy between the behavior of 
the spectral gap which occurs in a point ($d=0$) system and the
singularity in the free energy of a 
classical chain ($d=1$)  is an example of a deeper analogy
which relates quantum to thermal fluctuations; the 
formal correspondence $\hbar \leftrightarrow 1/(\beta \omega _{ 0})$
manifests a far-reaching analogy between $d$-dimensional 
quantum mechanics and $(d+1)$-dimensional classical statistical
mechanics. The analogy is most fruitful at $d=1$, because of the
interplay and the richness of exact available results which based either
in the transfer-matrix approach of 2-dimensional 
classical statistics or on the Bethe-Ansatz developed for
1-d quantum spin systems.}
 it is useful to examine the average 
displacement (\ref{OPTI}), determined by the ground-state (GS)
eigenfunction 
\begin{equation}
\phi _{ 0}(x) = e^{-\zeta /2 } \> \zeta ^{\delta -1/2 }
\label{MorseGSWF}
\end{equation}
where $\zeta = 2\delta e^{-ax }$. It is straightforward to see
that, as $T$ approaches $T_{ c}$ from below, the eigenfunction 
extends towards larger and larger positive values of $x$: 
\begin{equation}
\phi _{ 0}(x) \propto e^{-\lambda x } 
\end{equation}
where 
\begin{equation}
\lambda = \frac{ 1}{\delta -\delta _{ c} } 
\end{equation}
is a (transverse) characteristic length which measures the spatial 
extent of the GS eigenfunction. As a consequence, we can
estimate that  $<y>$, which is dominated by the large values of
the argument, will also behave as
\begin{equation}
<y> \sim ( \delta -\delta _{ c}   )^{-1 }\sim 
\left( 1-\frac{ T}{T_{ c} } \right)^{-1 }  \>.
\label{OPasy}
\end{equation}
As the critical temperature is approached from below,
particles cease to be confined to the minimum of the
Morse well. They perform larger and larger excursions
to the flatter part of the potential. At $T_{ c}$ the transition
is complete; the average transverse displacement is
infinite. Particles move, on the average, on the 
flat top of the Morse potential. Unwinding (``melting") 
of the DNA has occurred. \par
In the language of critical phenomena 
$<y>$ is the order parameter. In the ``usual" phase 
transitions, one goes from an ordered to a disordered
phase. The order parameter $m$ vanishes at the 
transition point, i.e $m \propto (T_{ c}-T)^{\beta  }$
with a positive critical exponent $\beta $ (not to be
confused with the inverse temperature: standard
notation of critical phenomena!). DNA melting is really \nopagebreak
an instability\cite{wetting}  - rather than an ``order-disorder" transition.
It is therefore  not surprising that the corresponding
critical exponent $\beta$ extracted from (\ref{OPasy})
is negative (-1).\par 
Experimental data on DNA denaturation do not deliver
$<y>$ directly. The ``experimental order parameter" is
the helical fraction, i.e. the probability that a given base
pair is still bound; technically one uses an
(instrumentation-dependent) cutoff $y_{ 0}$
and measures $P(y>y_{ 0},T)$. For the model presented
here, this function approaches zero smoothly (linearly) as
$T \to T_{ c}$, independently of the choice of $y_{0}$.

Eq. (\ref{xiTI}) states that the correlation length is also
contolled by the gap in the eigenvalue spectrum; as
the transition is approached,
\begin{equation}
\frac{ \xi }{l } = \frac{ 1}{\beta D } 
\left( 1 - \frac{ T}{T_{ c} }\right)^{-2 } 
\end{equation}
which identifies a critical exponent $\nu =2$ for the
divergence of the correlation length.  The picture of
thermal denaturation which emerges is one of ordered
regions, 
where helical structure persists; these regions
are interrupted by droplets of the high-temperature
phase, i.e. ``denaturation bubbles" of typical size $\xi $.  

\subsection{A first order transition?}
\label{subec:1storder}
It is possible to generalize the theory in order to
take account of the fact that the stacking energy is
a property of successive base pairs, rather
than individual bases. A practical way of doing this
is to substitute the second term in the Hamiltonian
(\ref{eq:trham}) by
\begin{equation}
\frac{1}{2}\mu \omega_{0}^{2}
\left[ 1 + g (y_{n} + y_{n-1} ) \right]
\left( y_{n} -  y_{n-1} \right)^{2}  \quad,
\label{eq:nlstack}
\end{equation}
where \cite{dauxpeyr1}
\begin{equation}
g(x)= e^{-\alpha x}   \quad.
\label{eq:expinterp}
\end{equation}
The effect of Eqs. (\ref{eq:nlstack})-(\ref{eq:expinterp}) is to interpolate
between the original value of the elastic coupling
if {\em either} (or both) of the two base pairs $n$, $n-1$ is
unbound (in which case $y_{n}\to \infty $  or $y_{n-1}\to \infty$),
and twice that value if both are bound; in the latter case, typically, $y_{n} \approx 0$;
the much higher values of the stacking energy, 
which (\ref{eq:expinterp}) in principle allows,
are statistically irrelevant due to the repulsive core of the Morse potential.
Within the gradient expansion approximation, it can be shown\cite{TDP}
that the main effect of the
nonlinear stacking energy on the thermodynamics is to generate an
effective, on-site, ``thermally activated" barrier
\begin{equation}
U(y) = \frac{T}{2D}\ln \left( 1 + e^{-2\alpha y } \right)  \quad.
\label{thbarrier}
\end{equation}
which appears in Eq.  \ref{Schr} and acts in addition to the Morse potential.

It has been shown\cite{dauxpeyr2,TDP,CuleHwa}  that the character of the
transition changes dramatically as the value of the
stacking parameter ratio $\alpha / a$ decreases (corresponding to
a longer range in the effective potential).
Although the transition remains asymptotically
second order, the limiting asymptotic behavior becomes relevant only
within an exponentially small range of the temperature difference
$T_{c}-T$. For all practical purposes, the transition is first order, with
a finite melting entropy $\Delta S =A_{0}D/T_{c}$, where $A _{0}$
is a numerical constant of order unity\cite{TDP}.  

It should be noted that the interpolation (\ref{eq:expinterp}) is not
unique; an interpolation function of the type $g(x) = 1, \> x<x_{0} $, 
 $g(x) = (x_{0}/x)^{2}, \> x>x_{0} $, leads - depending on the
other parameters - to a rich variety of critical behavior, ranging from a
first-order transition to continuously varying critical exponents.\cite{Zia}  

With the above modification (\ref{eq:nlstack}), 
it has become possible\cite{CuleHwa}
to describe, at least in principle, the 
series of multistep melting observed in real, heterogeneous DNA. 

\subsection{TI beyond the gradient expansion}
\label{subsec:TIbeyondgp}
It was stated in Section  \ref{subsubsec:TIdefnot} that the
TI formalism rests on the assumption that the integral equations (\ref{right})
and (\ref{left}) -  
have a complete, orthonormal set of eigenfunctions. Within the
gradient approximation approach this was demonstrated by
construction - since the integral equation was reduced to
a Schr\"odinger-like equation.  In many cases however, the
gradient expansion is not valid at all. It is therefore necessary to
develop alternative, mostly numerical methods for computing 
TI thermodynamics.  For such applications it is expedient to
consider the symmetrized, dimensionless version of the kernel (\ref{kernel}), i.e.
\begin{equation}
T_{s}(x,y) = e^{- \left[(y-x)^{2}/R +V(x)+V(y)  \right]  /(2T)} 
\label{symmkernel}
\end{equation}
and the associated integral equation
\begin{equation}
\int_{-\infty  }^{ \infty }dy T_{s}(x,y) \phi(y) = \Lambda \phi (x)
\quad,
\label{symmti}
\end{equation}
where $R=Da^2/(\mu \omega _{0}^{2 })$, $T=1/(\beta D)$ and
the Morse potential is now dimensionless, $V(x)=(1-e^{-x })^{2 }$, as
are the displacement variables $x,y$.  

Due to the flat top of the Morse potential, the kernel (\ref{symmkernel})
is not of the Hilbert-Schmidt type\cite{Zhang}; therefore the integral equation (\ref{symmti})
is singular and it can not be a priori stated that it possesses a complete
orthonormal set of eigenstates; in other words, the prerequisites for directly
applying the TI method are not  strictly met. In the rest of this section I will outline
a mathematically consistent procedure of examining the 
spectral gap of (\ref{symmti}),
based on finite-size scaling concepts\cite{nth_fss}. 

Due to the presence of the Gaussian factors
in the kernel, it is possible to approximate the integral in the left-hand-side of 
(\ref{symmti}) by using a  Gauss-Hermite grid of size $N$, i.e. 
\begin{equation}
\int_{ -\infty }^{ \infty } d{\bar y} \> e^{-{\bar y}^2 } f({\bar y}) 
\approx
\sum_{m=1}^{N}  w_{ m} f({\bar y}_{m})
\label{GH}
\end{equation}
where the positions $\{  {\bar y}_{ m}\}$ and weights $\{w_{m}\}$ 
are given by the appropriate
Gauss-Hermite quadratures routine. 
The largest ${\bar y}_{ N }\approx (2N+1)^{1/2}\equiv L$
can be used as estimate of the {\em transverse} ``size of the system" employed at any
given discretization. I emphasize transverse because the length of the chain is
infinite, i.e. the thermodynamic limit has already been taken. 
 
I use ``rescaled" variables, i.e.
$y=\rho {\bar y}$, $\rho =(2RT )^{1/2 }$,  divide
both sides of (\ref{symmti}) by $\rho  \sqrt{\pi }$, and use
the approximation (\ref{GH}). The result is an 
approximation of (\ref{symmti}) by the  
matrix eigenvalue equation 
\begin{equation}
\sum_{j=1}^{N} D_{ij} A_{ j}^{\nu  } =  {\tilde \Lambda_{ \nu } }A_{ i}^{\nu  }
\label{matrTI}
\end{equation}
where
\begin{equation}
D_{ij}= \left( \frac{ w_{ i} w_{ j} } {\pi  } \right)^{1/2} 
e^{ {\bar y}^{i }{\bar y}^{j } }
e^{ -( {\bar y}^{i } - {\bar y}^{j })^{2 }/2 }  
 e^{ -\left[  V(\rho {\bar y}^{i }) +V (\rho{\bar y}^{j } ) \right]/(2T) }
\label{Dmatr}
\end{equation}
and ${\tilde \Lambda_{ \nu } }\equiv \Lambda _{ \nu }/(2\pi R T )^{1/2}
\equiv e^{-\epsilon _{ \nu }/T }$. 
\begin{figure}[h]
\psfig{figure=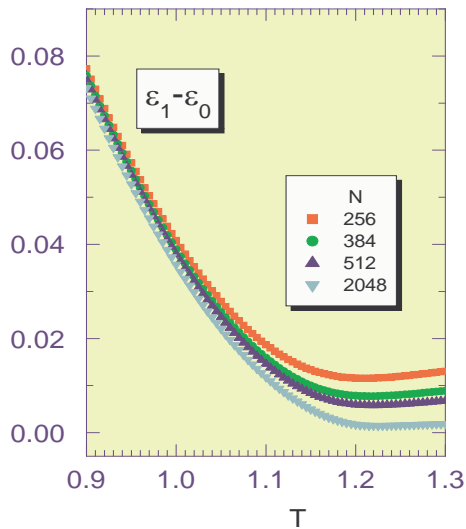,height=9truecm,width=7truecm}
\vskip  -.5truecm
\caption{The gap between the two lowest eigenvalues of the
matrix eigenvalue problem (\ref{matrTI}), for a variety
of $N$ values. For a given $N$, the gap has a minimum at
a certain temperature $T_{m}$. 
}
\label{Fgap}
\end{figure}
\begin{figure}[h]
\psfig{figure=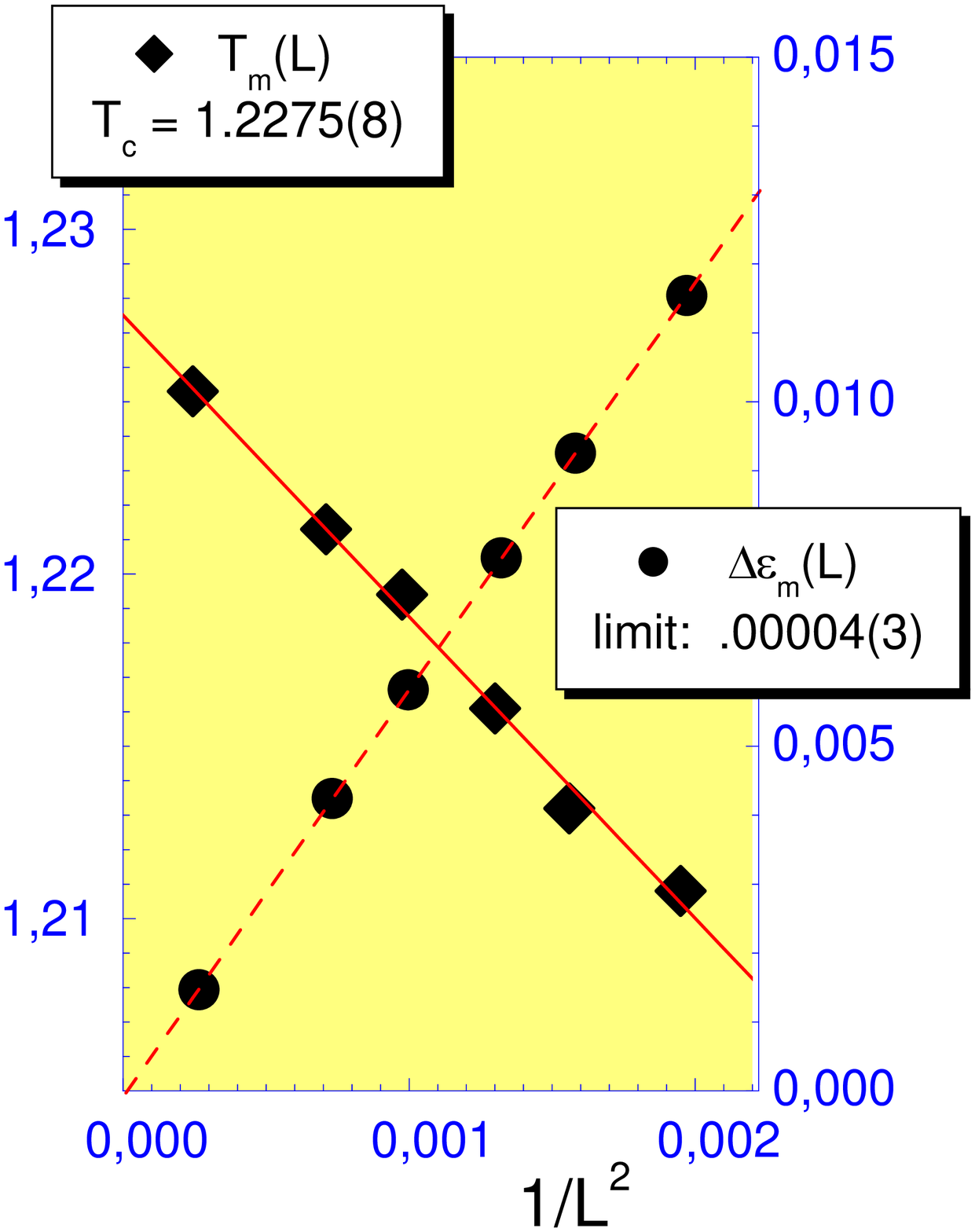,height=9truecm,width=7truecm}
\vskip -.5truecm
\caption{ The magnitude of the gap minimum (circles, right y-axis scale)
approaches 
zero as the system size goes to infinity.
The sequence of the temperatures corresponding
to the gap minima, $T_{m}(L)$ (diamonds, left y-axis scale),
can be used to provide an estimate of the  
critical point $T_{c}$. 
}
\label{TcLocate}
\end{figure}

It is now possible to solve numerically the real, symmetric matrix eigenvalue problem
(\ref{Dmatr}) for a range of temperatures and a sequence of increasingly
fine grids. 
Results for the difference between the two lowest eigenvalues are shown
in Fig. \ref{Fgap} for  $R=10.1$ .  For any given size $L$, the
gap has a minimum $\Delta \epsilon _{m}(L)$ at a certain temperature 
$T_{m}(L) $. Fig. \ref{TcLocate} 
demonstrates that (i) the value of the
gap approaches zero quadratically as  $L \to \infty $  (with an 
accuracy of  $10^{-5 })$,
and (ii) the sequence of $T_{m}(L)$'s also approaches a limiting value
$T_{c}=1.2275$  quadratically. 

It is natural to  identify the limiting temperature $T_{c}$, where the spectral gap of the
limiting, infinite-dimensional matrix eigenvalue equation (\ref{matrTI}) vanishes,
as the transition temperature of
the original TI equation (\ref{symmti}). Further application of finite-size scaling
methods demonstrates\cite{nth_fss} that the various critical exponents coincide with those
obtained within the gradient expansion method.

\section*{Acknowledgments}
I thank M. Peyrard and T. Dauxois for many helpful discussions and comments.
\section*{Appendix:
Phase
transitions in one-dimensional  systems}
\renewcommand{\theequation}{A.\arabic{equation}}
I briefly discuss why the general prohibitions on 
phase transitions in one dimension are inapplicable to
both the PS and the PB 
models of DNA denaturation.

Van Hove's theorem\cite{vanHove}
states that no phase transitions occur in 1-d particle systems
with short-range  pair interactions. The PB 
model has on-site potential - i.e. the theorem is not applicable.
It is however worth noting that similar mathematical proofs, 
have been given for systems
with periodic on-site potentials\cite{CueSa}. Such proofs however 
seek to prove analyticity of the eigenvalue
spectrum and hence absence of a phase transition; as such, they 
tend to exclude potentials which give rise
to singular TI equations. 

The PS model is not a Hamiltonian model and therefore van Hove's theorem is
again not applicable. 

Landau's theorem\cite{Landau} is significantly stronger. It
states that ``macroscopic phase coexistence cannot occur at finite temperatures
in one dimensional systems".  It is less obvious why it should not apply. 
I therefore outline the proof. 
Consider a system with $N$ sites, which may exist in either
phase $A$ or phase $B$. Let $\theta $ be the fraction of
phase $A$; furthermore, let there be $m<<N$ contacts between
the phases, each of energy $\epsilon $. These can be 
steplike (Ising) or continuous domain walls. The free
energy of the configuration is given by
\begin{equation}
F=N\theta f_{A} + N(1-\theta) f_{B} + F_{DW}
\label{totfe}
\end{equation}
where
\begin{equation}
F_{DW}= m \epsilon - k_{B}T S_{DW}(m,N)
\end{equation}
and the (dimensionless) entropy is given by
\begin{equation}
S_{DW}(m,N) = \ln\left[\frac{N!}{m!(N-m)!}\right]
\approx  m \ln \left[\frac{Ne}{m}\right]  \quad.
\end{equation}
Minimization of the total free energy with respect to $m$
yields
a macroscopic average number (ie. a \emph{finite density}) 
of domain walls
\begin{equation}
{\bar m}= N e^{-\epsilon/(k_{B}T) }  \>.
\end{equation}
The system breaks up into $m$
regions of finite size $e^{\epsilon/(k_{B}T) }$. Macroscopic
phase separation can only occur at zero temperature 
(as the domain size goes to infinity).

Landau's argument 
covers a wide range of systems, e.g. double-well on-site
potentials (Ising universality class), or periodic on-site potentials. 
It does not cover the PB case, because the DW has infinite energy\cite{DTP}.
It does not apply to the PS case because the loop entropy is
not proportional  to the size of the loop and therefore (\ref{totfe}) 
does not hold. On the contrary, the theorem is applicable to the generalized
zipper model, as its authors had correctly noted\cite{ZimmBragg}.

\end{document}